\documentclass[twocolumn]{aastex631}

\newcommand{\porb}{$P_{orb}$}
\newcommand{\pspin}{$P_{spin}$}

\newcommand{\suzaku}{{\it Suzaku}}
\newcommand{\xmm}{{\it XMM}-Newton}

\newcommand{\swift}{{\it Swift}}

\newcommand{\sgr}{V1082\,Sgr}
\usepackage{xcolor}
\usepackage{soul}

\shorttitle{Detection of the white-dwarf spin of V1082 Sgr}
\shortauthors{Lima et al.}
\listofchanges
\graphicspath{{./}{figures/}}
\begin{document}

\title{Detection of the white-dwarf spin of the long-orbital period magnetic cataclysmic variable V1082~Sgr}

\author[0000-0001-6013-1772]{I. J. Lima}
\affiliation{Universidade Estadual Paulista ``J\'{u}lio de Mesquita Filho'', UNESP, Campus of Guaratinguet\'{a}, Av. Dr. Ariberto Pereira da Cunha, 333 - Pedregulho, Guaratinguet\'{a} - SP, 12516-410, Brazil\\}
\affiliation{CONICET-Universidad de Buenos Aires, Instituto de Astronom\'{i}a y F\'{i}sica del Espacio (IAFE), Av. Inte. G\"{u}iraldes 2620, C1428ZAA, Buenos Aires, Argentina\\} 
\affiliation{Universidad Nacional de San Juan, Facultad de Ciencias Exactas, Físicas y Naturales, Av. Ignacio de la Roza 590 (O), Complejo Universitario "Islas Malvinas", Rivadavia, J5402DCS, San Juan, Argentina\\}

\author[0000-0002-2953-7528]{G. Tovmassian}
\affiliation{Institute of Astronomy, Universidad Nacional Autonoma de Mexico, Ensenada, Baja California\\}
\affiliation{INAF – Osservatorio Astronomico di Brera, Via E. Bianchi 46, 23807 Merate (LC), Italy\\}
%\collaboration{6}{(AAS Journals Data Editors)}

\author[0000-0002-9459-043X]{C. V. Rodrigues}
\affiliation{Instituto Nacional de Pesquisas Espaciais (INPE/MCTI), Av. dos Astronautas, 1758, S\~ao Jos\'e dos Campos, SP, Brazil\\}

\author[0000-0001-6422-9486]{A. S. Oliveira}
%\altaffiliation{AASTeX v6+ programmer}
\affiliation{IP\&D, Universidade do Vale do Para\'\i ba, 12244-000, S\~ao Jos\'e dos Campos, SP, Brazil\\}

\author[0000-0002-2647-4373]{G. J. M. Luna}
\affiliation{Universidad Nacional de Hurlingham (UNAHUR). Secretaría de Investigación, Av. Gdor. Vergara 2222, Villa Tesei, Buenos Aires, Argentina\\}
\affiliation{Consejo Nacional de Investigaciones Científicas y Técnicas (CONICET)\\}

\author[0000-0002-7004-9956]{D. A. H. Buckley}
\affiliation{South African Astronomical Observatory, PO Box 9, Observatory 7935, Cape Town, South Africa\\}
\affiliation{Department of Physics, University of the Free State, PO Box 339, Bloemfontein 9300, South Africa\\}
\affiliation{Department of Astronomy, University of Cape Town, Private Bag X3, Rondebosch 7701, South Africa\\}

\author[0000-0003-1949-4621]{K. M. G. Silva}
\affiliation{Gemini Observatory/NSF's NOIRLab, Casilla 603, La Serena, Chile\\}

\author[0000-0002-0337-1363]{A. C. Mattiuci}
\affiliation{Instituto Nacional de Pesquisas Espaciais (INPE/MCTI), Av. dos Astronautas, 1758, S\~ao Jos\'e dos Campos, SP, Brazil\\}

\author[0000-0001-9255-864X]{D. C. Souza}
\affiliation{IP\&D, Universidade do Vale do Para\'\i ba, 12244-000, S\~ao Jos\'e dos Campos, SP, Brazil\\}

\author[0000-0002-7095-4147]{W. Schlindwein}
\affiliation{Instituto Nacional de Pesquisas Espaciais (INPE/MCTI), Av. dos Astronautas, 1758, S\~ao Jos\'e dos Campos, SP, Brazil\\}

\author[0000-0002-8646-218X]{F. Falkenberg}
\affiliation{Instituto Nacional de Pesquisas Espaciais (INPE/MCTI), Av. dos Astronautas, 1758, S\~ao Jos\'e dos Campos, SP, Brazil\\}

\author[0000-0002-0396-8725]{M. S. Palhares}
\affiliation{IP\&D, Universidade do Vale do Para\'\i ba, 12244-000, S\~ao Jos\'e dos Campos, SP, Brazil\\}



\begin{abstract}
We report on the discovery of circular polarization modulated with a period of 1.943~$\pm$~0.002~h in the cataclysmic variable V1082 Sgr. 
These findings unambiguously reveal the rotation of a magnetic white dwarf and establish its intermediate polar (IP) nature. Along with its extraordinary long orbital period (\porb) of 20.8~h, the spin period (\pspin) places this system in an extreme position of the \pspin\ versus \porb\ distribution. The circular polarization phase diagram has a single peak and an amplitude smaller than 1\%. These data were used to model the post-shock region of the accretion flow on the white-dwarf surface using the CYCLOPS code. We obtained a magnetic field in the white-dwarf pole of 11~MG and a magnetospheric radius consistent with the coupling region at around 2 -- 3 white-dwarf radii. 
The \pspin /\porb\ value and the estimated magnetic field momentum suggest that \sgr\ could be out of spin equilibrium, in a spin-up state, possibly in a stream accretion mode.

\end{abstract}



\section{Introduction} \label{sec:intro}

\sgr\ is a cataclysmic variable (CV) with an unusually long orbital period, \porb, of 20.82~h and a complex photometric variability \citep{Cieslinski_1998, Thorstensen_2010, Bernardini_2013, 2016ApJ...819...75T}. \porb\ was determined from radial velocity modulation of the absorption lines from a K-type secondary star \citep{Thorstensen_2010,2016ApJ...819...75T,Tovmassian_2018b}. 
\porb\ is also detected in photometry when the system is in the deep low-state \citep[using \textit{Kepler K2} data]{Tovmassian_2018a}.
The system features strong single-peaked emission lines of hydrogen and helium, except when it is in a short-lived very low state, when no emission line is detected. In the last years, mounting observational evidence points to the presence of a magnetic white dwarf (WD). 
A strong He~II~$\lambda$4686 emission line and hard X-ray emission, typical of magnetic CVs, have been reported by \citet{Cieslinski_1998}, \citet{Thorstensen_2010} and \citet{Tueller_2010}.  
In this regard, \citet{Bernardini_2013} detected a transient modulation with a time-scale of about 2~h in the \xmm\ X-ray light curve. Given its transient nature, they discarded its relation with the WD spin, and associated it to the mass transfer mechanism. Their X-ray spectrum is consistent with a multitemperature optically thin thermal plasma, often found in magnetic CVs. The object is highly variable in optical, UV, and X-ray bands, mostly irregular, showing some unstable periodic or cyclical variations on different time scales.
A quasi-periodic photometric variability of about 29 days, probably associated with variations in the accretion rate,  was detected by \citet{2016ApJ...819...75T, Tovmassian_2018a}.

There are two main classes of magnetic CVs. Polars have WDs that rotate (\pspin ) synchronously or near-synchronously with \porb\ and Roche-Lobe filling secondary star's rotation \citep{Cropper_1990}, whereas the intermediate polars (IPs) have asynchronous WDs and most of them have \pspin /\porb ~$\approx$~0.01~--~0.6 \citep{Patterson_1994}. The IPs evolve to shorter \porb\ hence its spin-orbital period ratio is expected to increase (e.g., \citealt{Norton_2008}). Two subtypes of polars are the prepolars and Low Accretion Rate Polars (LARPs) \citep{Schwope_2002, Schwope_2009}. Both of them are synchronous systems. Historically, some objects now understood as prepolars were believed to be LARPs. Presently, these two types are considered to represent different evolutionary stages of a magnetic CV. LARPs are polars that undergo prolonged low-accretion states in temporarily detached systems (e.g., \citealt{ferrario/2015}). On the other hand, prepolars are short \porb\ systems (i.e., \porb\ $< 6$\ h) in which a M type secondary star does not yet fill its Roche lobe, but can show some accretion fed by the secondary's wind.
Given recent discoveries of detached AR~Sco-like systems, with magnetic WDs, prepolars can eventually evolve into polars (e.g., \citealt{Schreiber_2021}).

In a seminal work, \cite{1977ApJ...212L.125T} found that the linear polarization of AM~Her, the polar prototype, varies with its \porb. Since then, many other polars showing variable linear and/or circular polarization were identified. 
The polarized emission from a magnetic CV is originated by cyclotron radiation produced in the so-called post-shock region (PSR) and the variable polarization is the result of a changing view of the PSR \citep{1977ApJ...218L..57S}. Cyclotron emission can also be established from the detection of cyclotron humps in the spectral continuum or spectropolarimetry. Polarization is usually high in polars (tens of percent). Some IPs \citep[e.g.,][]{Butters_2009} and prepolars \citep[e.g.,][]{2022MNRAS.513.3858H} exhibit small values of polarization, which is interpreted as a confirmation of magnetic accretion.

The exact classification of V1082\,Sgr has not yet been settled. It was initially classified as a symbiotic system \citep{Cieslinski_1998}, while \citet{Thorstensen_2010} and \citet{Bernardini_2013} argue in favor of a magnetic CV interpretation. \citet{2016ApJ...819...75T,Tovmassian_2018a, Tovmassian_2018b} consider the classification as a detached system: it would be one of the first long-period prepolars, i.e., a prepolar with an early-K companion. However, \citet{2019MNRAS.489.3031X} proposed an evolutionary model in which V1082\,Sgr is an IP with a Roche Lobe filling K-type secondary.  

No previous study has detected polarization in \sgr\ that would be an unquestionable evidence of a magnetic WD in the system and useful to disentangle its classification. Here, we report on the results of circular photo-polarimetry of V1082\,Sgr. We detected circular polarization in this system modulated with a period of 1.943~h, which we interpret as the spin period, \pspin, of the WD, characterizing it as a long-period IP. 
Section~\ref{observation} describes the observations and data reduction.
The results and the data analysis are shown in Section~\ref{data_analysis}. Section~\ref{sec:cyclops_model} presents the modeling of the cyclotron emission of V1082\,Sgr. The discussion and conclusions are presented in Section~\ref{discussion} and Section~\ref{sec:conclusions}, respectively.

\section{Observation and data reduction} \label{observation}

V1082~Sgr was observed in 10 nights spread over August and September 2020 using the 0.6~m Boller and Chivens (B\&C) telescope of the Observat\'orio do Pico dos Dias (OPD) operated by the Laboratório Nacional de Astrofísica (LNA), Brazil. Table~\ref{tab_data} presents a summary of these observations. The polarimeter {\sc IAGPOL} \citep{Magalhaes_1996, rodrigues1998} was configured with a quarter-wave retarder plate and a Savart plate. The observations were carried out using the CCD Ixon~4335, which has a pixel size of 13.5~$\times$~13.5~$\mu$m. The plate scale is 0.34"/pixel in the above instrumental setup. 

All datasets were obtained in white light (i.e. no filters were used) in order to maximize the count rates. As pointed out by \cite{Bruch_2017}, an estimate of the eﬀective wavelength of the white light bandpass is 5530~\AA, which is close to the effective wavelength of the Johnson V band (5500~\AA). We therefore calibrate the V1082\,Sgr magnitude using the star NOMAD~0692-0848091 ($\alpha_{2000.0}$=~19$^{h}$:07$^{m}$:17.17$^{s}$ and $\delta_{2000.0}$~=~-20$^{\circ}$:47$^{'}$:33.02$^{"}$), which has 13.9~mag in V band \citep{Zacharias_2005}. 
\sgr\ has an average V magnitude of 14.7~mag in {\sc NOMAD} catalog. 

The data reduction was performed using bias frames and dome calibration flat-field images to account for the detector's zero noise level and sensitivity. We also used polarized and unpolarized standard stars measurements to correct the linear polarization angle to the equatorial reference system and to verify the presence of spurious instrumental polarization. All procedures used the IRAF\footnote{IRAF (Image Reduction and Analysis Facility) is distributed by the National Optical Astronomy, which is operated by the Association of Universities for Research in Astronomy, Inc., under a cooperative agreement with the National Science Foundation \citep{Tody/1986, Tody/1993}.} and the {\sc PCCDPACK\_INPE}\footnote{{\sc PCCDPACK\_INPE} can be downloaded from \url{https://github.com/claudiavr/pccdpack_inpe/}} \citep{2018Pereyra, pereyra/2000} packages. Tables~\ref{pol_stard} and~\ref{unpol_stard} show the linear and circular polarization values of six standard stars from the southern hemisphere, observed on the same nights as V1082~Sgr. We confirmed that the instrumental circular polarization is negligible based on those values.
The polarization was calculated from the ordinary and extraordinary ray fluxes ratio as described in \cite{rodrigues1998}. A correction for different efficiencies from ordinary and extraordinary counts was applied using a normalization factor as described by \cite{Lima_2021}. Each circular polarization point is calculated using differential flux measurements of a moving set of 8 images gradually displaced across the time series. The circular polarization signal is not calibrated, i.e., the real signal of the circular polarization can be the inverse of the one presented here.

\begin{deluxetable*}{cccc}
\tablecaption{Summary of the observations.\label{tab_data}}
%\tablewidth{0pt}
\tablehead{
\colhead{Date Obs.} & \colhead{Exp. time} & \colhead{Time span}  &
\colhead{Mean mag.} \\
& \colhead{(s)} & \colhead{(h)} & \colhead{(mag)} 
}
%\decimalcolnumbers
\startdata
2020 Aug 22 & 200 & 4.5 & 14.0~$\pm$~0.3\\
2020 Aug 23 & 120 & 6.2 & 14.2~$\pm$~0.2\\
2020 Aug 24 & 120 & 5.6 & 14.1~$\pm$~0.1\\
2020 Aug 25 & 120 & 5.9 & 14.2~$\pm$~0.1\\
2020 Aug 26 & 100 & 6.1 & 14.1~$\pm$~0.1\\
2020 Aug 27 & 130 & 2.8 & 13.9~$\pm$~0.1\\
2020 Aug 30 & 150 & 6.1 & 14.2~$\pm$~0.2\\
2020 Aug 31 & 150 & 5.7 & 13.7~$\pm$~0.2\\
2020 Sep 01 & 150 & 2.6 & 14.2~$\pm$~0.1\\
2020 Sep 02 & 150 & 6.1 & 14.5~$\pm$~0.1\\
\enddata
%\tablecomments{}
\end{deluxetable*}

\section{Results and data analysis} \label{data_analysis}

\subsection{Photometry} \label{photometry}

\begin{figure*}
\centering
\includegraphics[width=\textwidth]{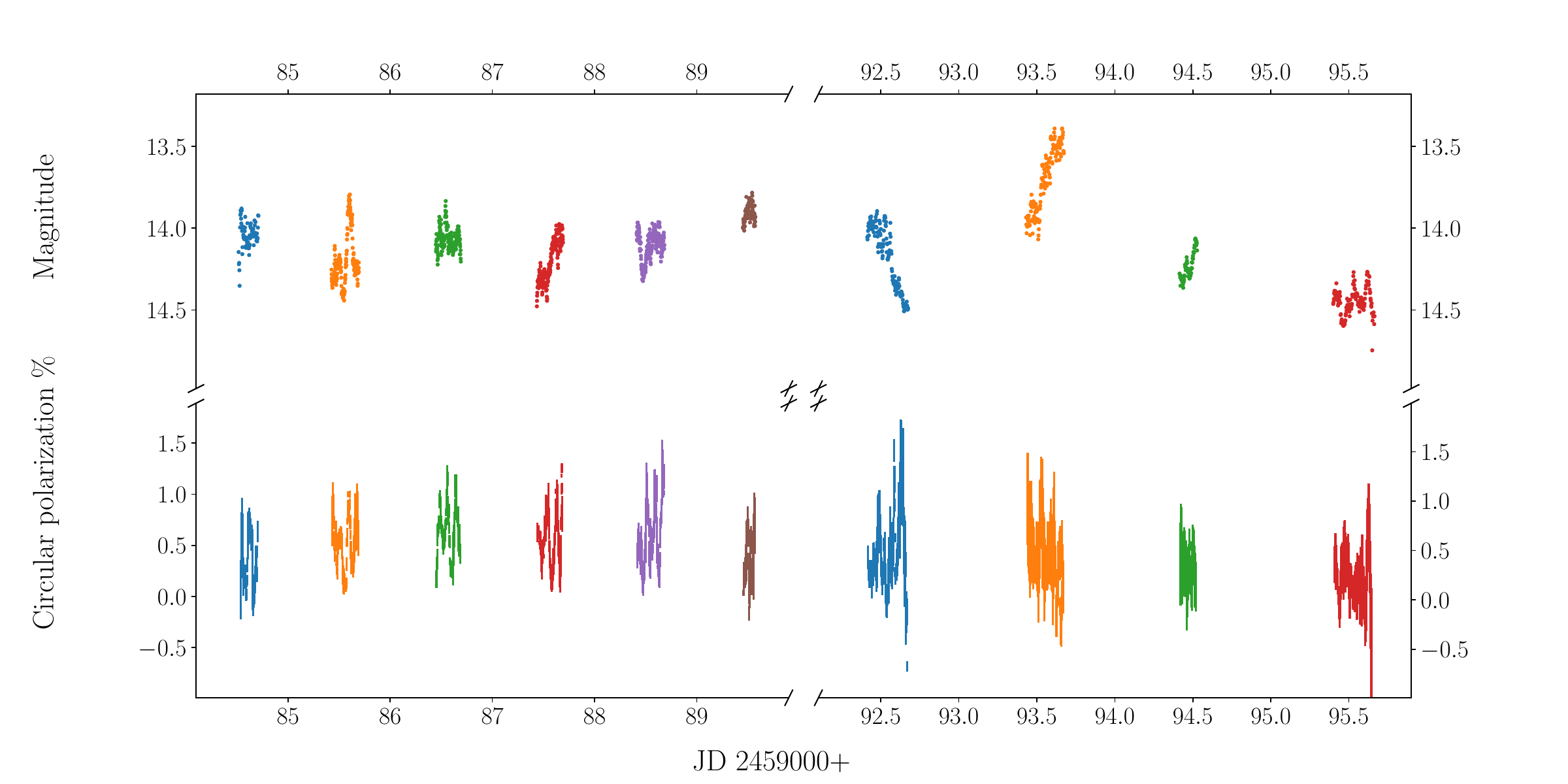}
\caption{Photometry and circular polarimetry of V1082~Sgr over ten nights of observations. The measurements were taken in white band. The magnitude values were calibrated to the V band, as explained in the text.}
\label{fig:LC}
\end{figure*}

The light curve of V1082~Sgr is shown in the upper panel of Figure~\ref{fig:LC}. The system shows an average magnitude of 14.1~mag in the clear band calibrated to V band as explained in the previous section. 
Table~\ref{tab_data} shows the average magnitudes for each night. The photometric behavior of the system changed after the gap: the system shows a rapid brightness decline, then the system reaches its maximum flux in our observations, and then it evolves in a steady decline in the last nights (see Figure\,\ref{fig:LC}). 

Our photometric data show no definite periods. The precise \porb\ of the system (20.82~h) was determined from the spectroscopy \citep{Thorstensen_2010, Tovmassian_2018b}. \porb\ was also detected in the photometric time series obtained by K2 mission \citep{Tovmassian_2018a}, but only at a very low luminosity state during small fractions of the time. In the same 80~days long continuous monitoring of the system brightness by K2, there was no clear \pspin\ signal. \citet{Tovmassian_2018a} noted a relatively significant peak at 11.2 cycles/day in the power spectrum, proposing it might be the optical counterpart of the frequency detected by \citet{Bernardini_2013}. 

\subsection{Polarimetry} \label{circular_polarimetry}

The polarimetric observations of V1082\,Sgr were obtained over ten nights, which were divided into two blocks -- Block~A (first six nights) and Block~B (last four nights) -- separated by a gap of two nights. The small angular distance to the moon during observations in Block~B caused the background to be higher, therefore Block~A data have higher quality than Block~B.

\begin{figure}
\centering
\includegraphics[width=\columnwidth]{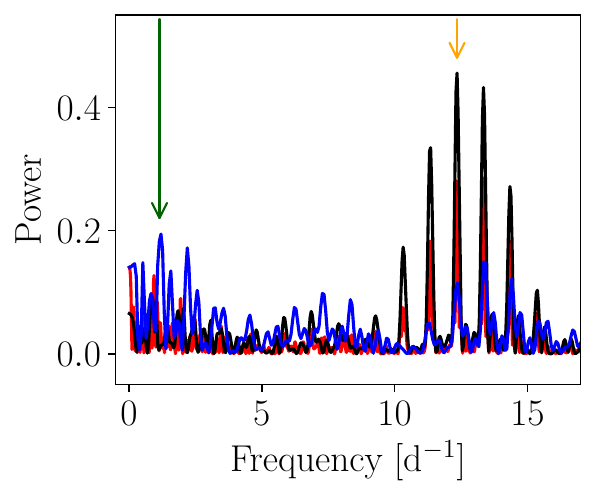}
\caption{Power spectra of the circular polarization curves before de-trending. The black line corresponds to the LS periodogram of Block~A; the blue is for Block~B, and the red corresponds to the entire data set. The vertical green arrow indicates the orbital frequency only detected in Block~B. The spin frequency is marked with an orange arrow.}
\label{fig:powpol}
\end{figure}

Our polarimetry is shown in the lower panel of Figure~\ref{fig:LC}. The circular polarization degree measurements of V1082~Sgr are clearly variable in the range from -0.5\% to 1.5\%, but the individual points have larger errors. To search for periodicities, we calculated the power spectra of the circular polarization data using the Lomb-Scargle (LS) \citep{Lomb_1976,Scargle_1982} procedure, which was performed for the entire data set as well as separately for Blocks~A and B (Figure\,\ref{fig:powpol}). The power spectrum of Block~A data shows a dominant peak at a frequency of 12.352~cycles~d$^{-1}$, or a period of 0.081~d. We interpret this as the spin frequency of the WD in V1082~Sgr.
In the case of Block~B, there is a series of 1-day aliases peaks around $\sim$13~cycles~d$^{-1}$ (in both blocks, data were obtained nightly), consistent with the strongest peak found in Block~A, while the highest power is located at the frequency of 1.154~cycles~d$^{-1}$, corresponding to the orbital movement. 

\begin{figure*}
\centering
\includegraphics[width=0.8\textwidth]{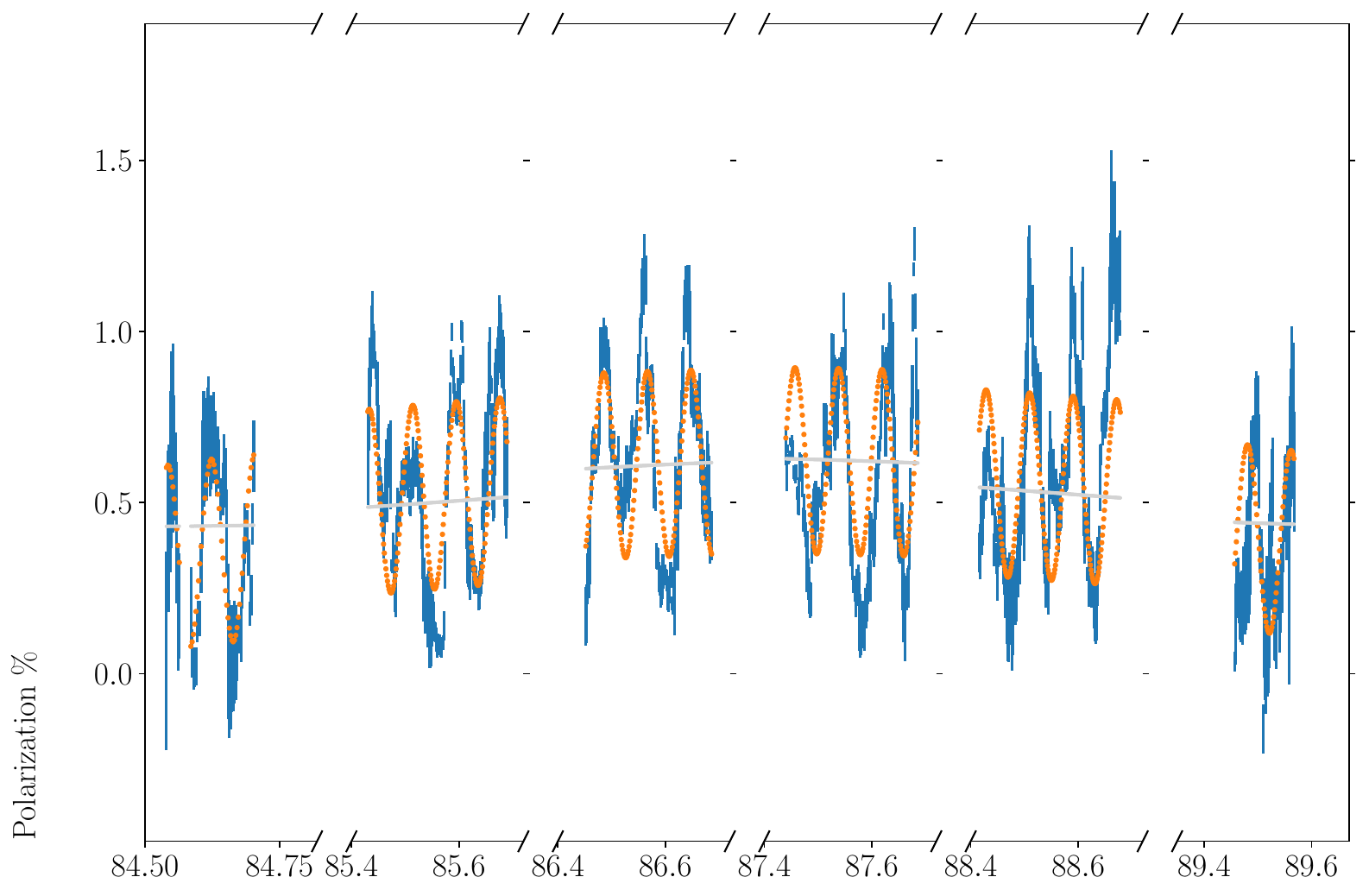}
\includegraphics[width=0.8\textwidth]{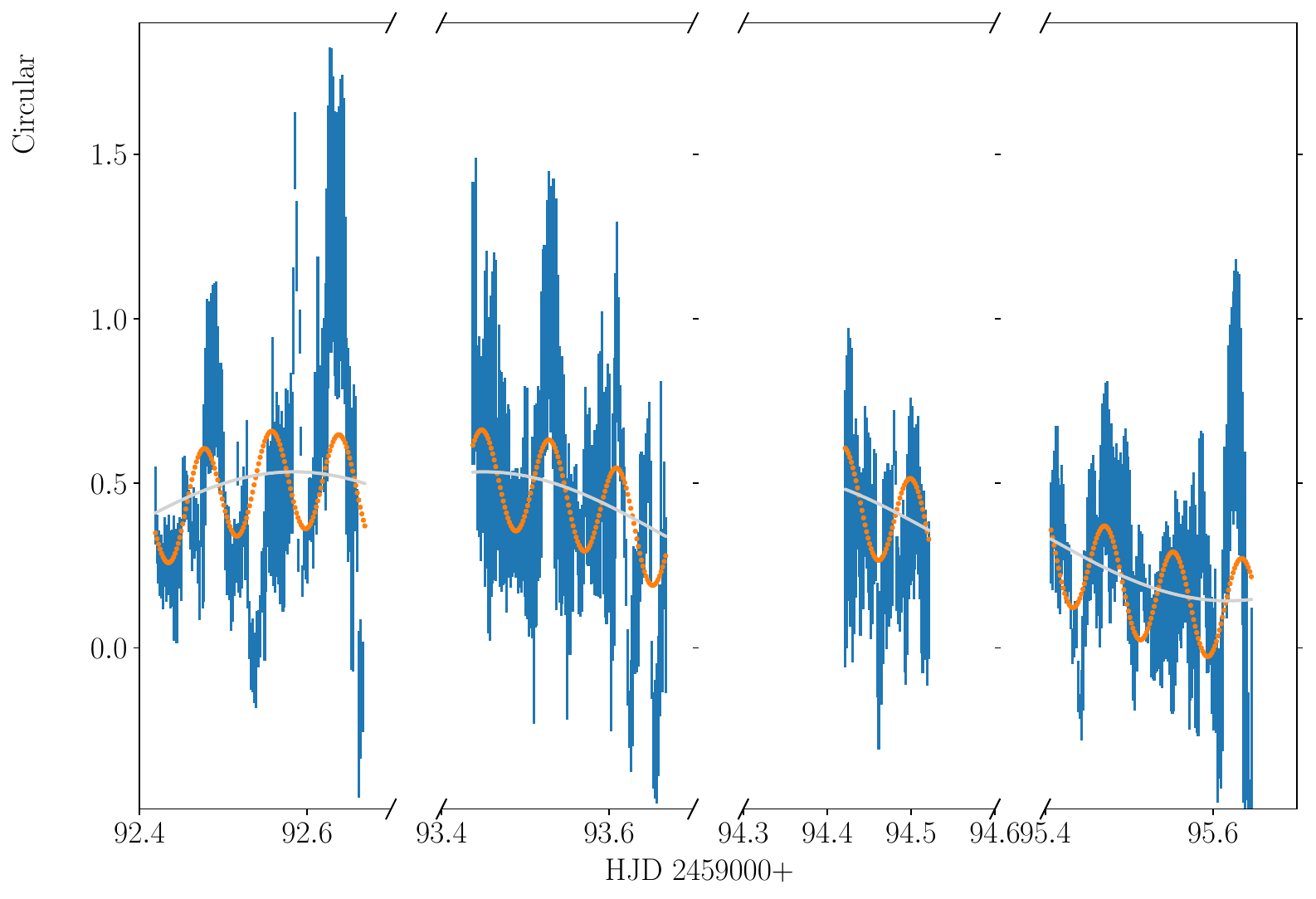}
\caption{V1082~Sgr circular polarization observations for Block~A (top panel) and Block~B (lower panel). The blue line represents the data, with error bars, and the orange line is the fit with \pspin\ + f$_{2}$, where f$_{2}$ is a low-frequency curve. In Block~A, it roughly corresponds to the time length of Block~A, and in Block~B, f$_{2}$= \porb. The f$_{2}$ curve is presented as a gray line.}
\label{fig:PCL}
\end{figure*}

In Figure~\ref{fig:PCL}, we show the circular polarization degree curves for Blocks~A (top) and B (bottom). 
The curves present a fast modulation ($\sim$0.08~d) superposed on long-period trends. This $\sim$0.08~d period is obviously related to the peaks seen in the power spectra. The longer trends were modeled by sinusoidal functions whose frequencies were defined by the main low-frequency peak in each power spectrum, i.e. f$_{2}$=0.185~cycles~d$^{-1}$ for Block~A (which roughly corresponds to the length of the block) and f$_{2}$=1.154~cycles~d$^{-1}$ for Block~B. The sinusoidal fits (gray lines in Figure~\ref{fig:PCL}) were then subtracted to detrend the data.

The LS power spectra of the detrended data show the same peaks as the original data, specially the peak at frequency 12.352~cycles~d$^{-1}$, or 1.943~h. We supplemented the time series periodicity analysis with other methods such as the Generalized Lomb-Scargle (GLS), the Discrete Fourier Transform (DFT), the Date Compensated Discrete Fourier Transform (DCDFT), and the Phase Dispersion Minimization (PDM), all available in the software \mbox{{\it PERANSO}} \citep{Paunzen_2016}. The periods obtained are all consistent with each other, pointing to P=0.08096~$\pm$~0.00005~d (=1.943~$\pm$~0.002~h), where the error is the standard deviation of the estimates using the different methods.

\begin{figure}[ht!]
\includegraphics[width=\columnwidth]{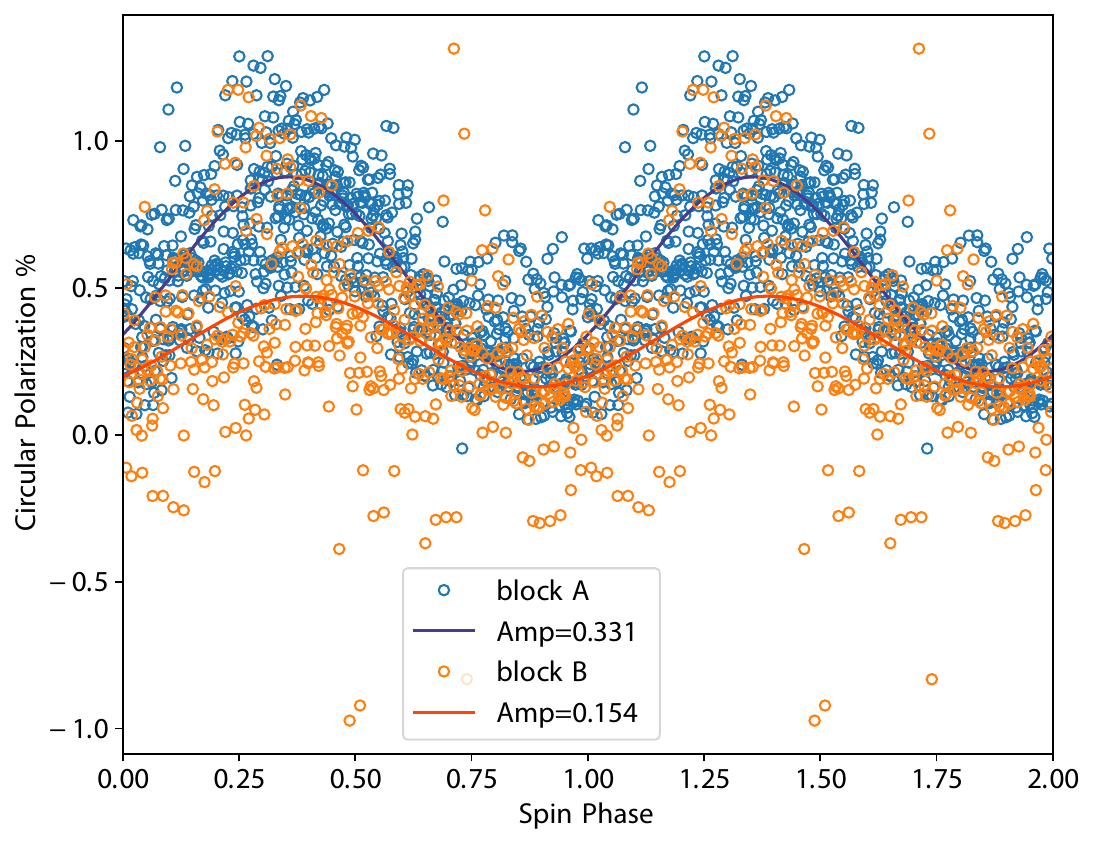}
\caption{The circular polarization degree folded with the 1.943\,h \pspin\ of the magnetic WD in V1082\,Sgr. Data in Blocks~A and B are represented by blue and orange points, respectively. The values have been subtracted from long-term trends as explained in the text. The amplitudes are shown in the box. Phase zero corresponds to the spectroscopic orbital T$_0$.\label{fig:phase}}
\end{figure}

The phase diagram of the detrended circular polarization degree, folded at the 1.943~h period and adopting T$_0$ from the spectroscopic stellar conjunction \citep{Tovmassian_2018b}, is displayed in Figure~\ref{fig:phase}, for each block. Both curves present one maximum and one minimum per cycle, and do not show changes in signal (the best-fitting sinusoids are always positive). However, there are differences in the amplitudes between the blocks, the amplitude in Block~A being about twice that of Block B.

To explore the presence of \porb\ in the circular polarization degree data from Block~B. The polarization degree is the ratio between the polarized flux and the total flux, so it is affected by changes in the total flux that can come from different parts of the system. The circularly polarized flux, on the other hand, is an emission purely from the PSR. 
Within the uncertainties, the results of the timing analysis of the polarized flux and polarization degree in Block~A are essentially the same. Block~B data, however, do not show the \porb\ modulation in the polarized flux, so we conclude that it is due to the total flux and it does not originate from the PSR. The \pspin\ is present in the polarized flux of Block~B, but with a smaller power than in Block~A. Therefore, the consistent presence of the 1.943~h period across both data segments, as well as in both the degree of polarization and the polarized flux, confirms that it represents the rotational period of a magnetic white dwarf.

\section{Modeling the cyclotron emission of V1082~Sgr}
\label{sec:cyclops_model}

The polarized emission from magnetic CVs is due to the cyclotron radiation from the PSR. This region is located in the end of the magnetic accretion column near the magnetic poles of the WD. It is formed by the compression caused by a shock that occurs just above the WD surface. 
To estimate some properties of the PSR in \sgr, we performed a modeling of its circularly polarized flux.
For that, we used the {\sc cyclops} code, which solves the radiative transfer in a 3D PSR 
\citep{Belloni_2021_cyclops,Silva_2013,Costa_2009}. The temperature and density profiles are consistently calculated considering the physical and geometric properties of the system as described in \cite{Belloni_2021_cyclops}. Some model parameters are the WD mass, the mass accretion rate, and the WD magnetic field, to cite the most relevant ones. 

In the modeling, we use only Block~A data. The input values to the model are circular polarized fluxes, which were obtained multiplying the observed flux (in arbitrary units) by the circular polarization degree. No detrend was applied to the data. 
The result was then folded in 50 bins of spin phases. The total flux of \sgr\ is very variable, and it does not present any evidence of modulation with the \pspin, as found in the polarization. Therefore, we chose not to model the total flux since the contribution from the PSR should be negligible compared with the emission from other system structures.
On the other hand, the polarized flux comes only from the PSR and is a clean signal from this region.
The emission was assumed monochromatic and calculated for a wavelength of 500~nm. 

\begin{figure}[tbh]
%\centering
% \includegraphics[width=9.0 cm,trim=10 0 80 0]{cyclops_model_x2019.pdf}
% \includegraphics[width=\columnwidth]{stokes_maps_x2019.pdf}
% \includegraphics[width=4.5 cm, trim=150 70 140 0]{psr_00_x2019.pdf}\includegraphics[width=4.5 cm, trim=150 70 140 0]{psr_05_x2019.pdf}
\includegraphics[width=1.1\columnwidth]{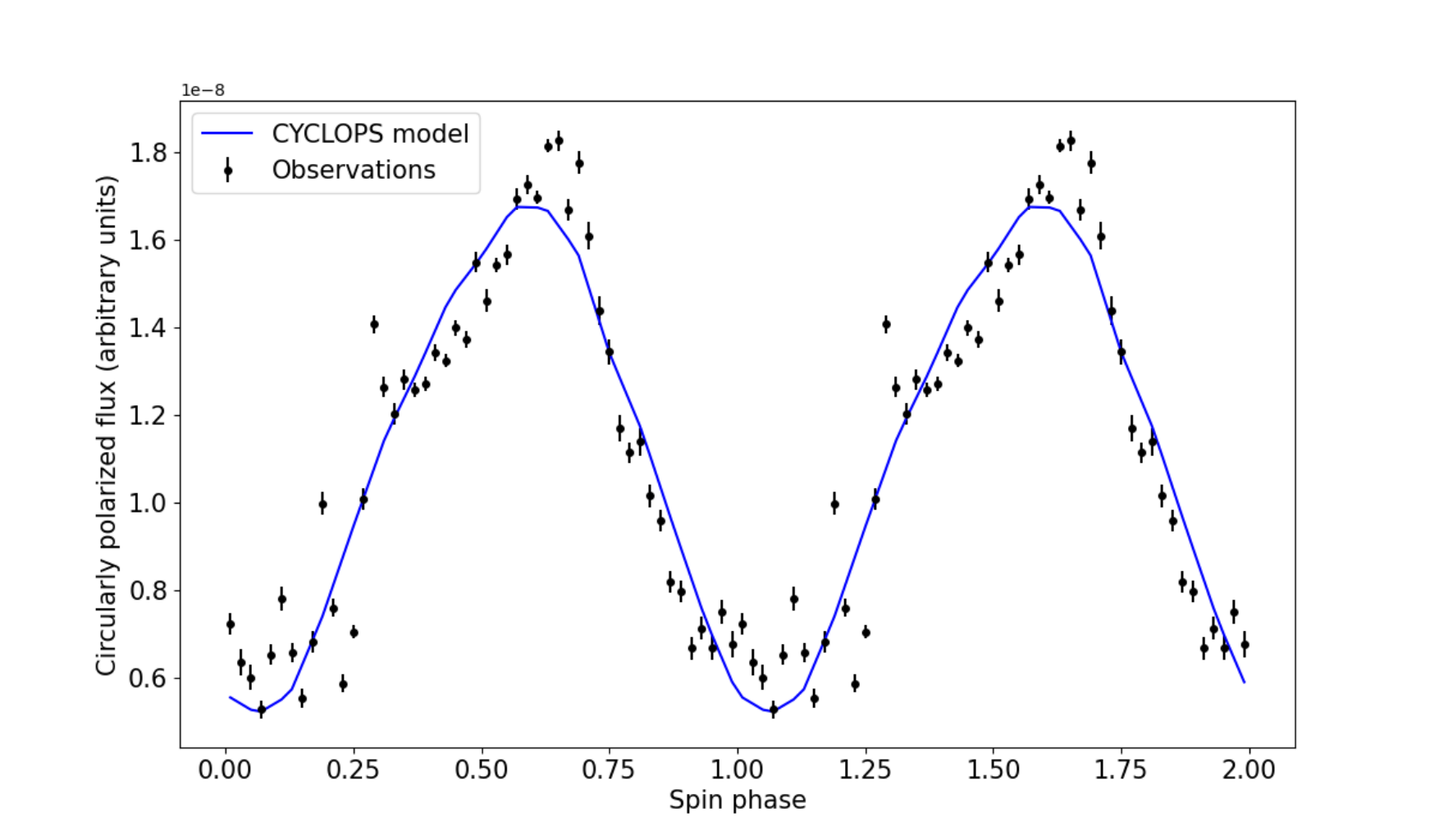}
\includegraphics[width=\columnwidth]{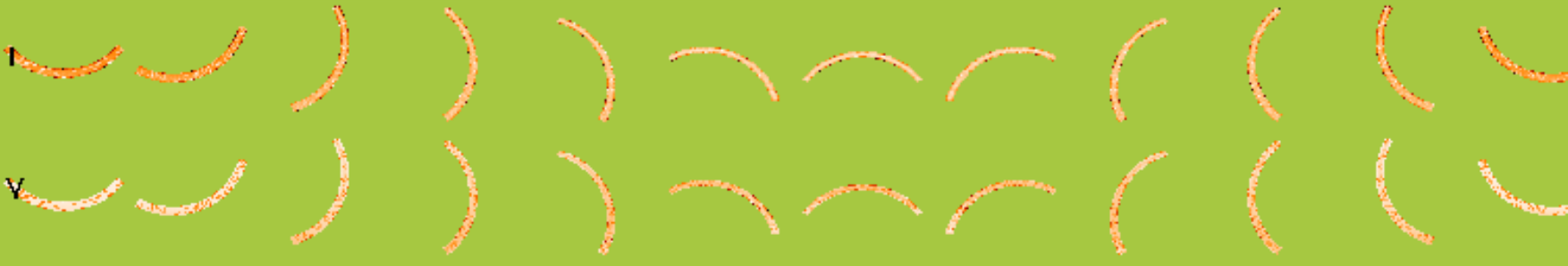}
\includegraphics[width=4.3 cm, trim=150 70 140 0]{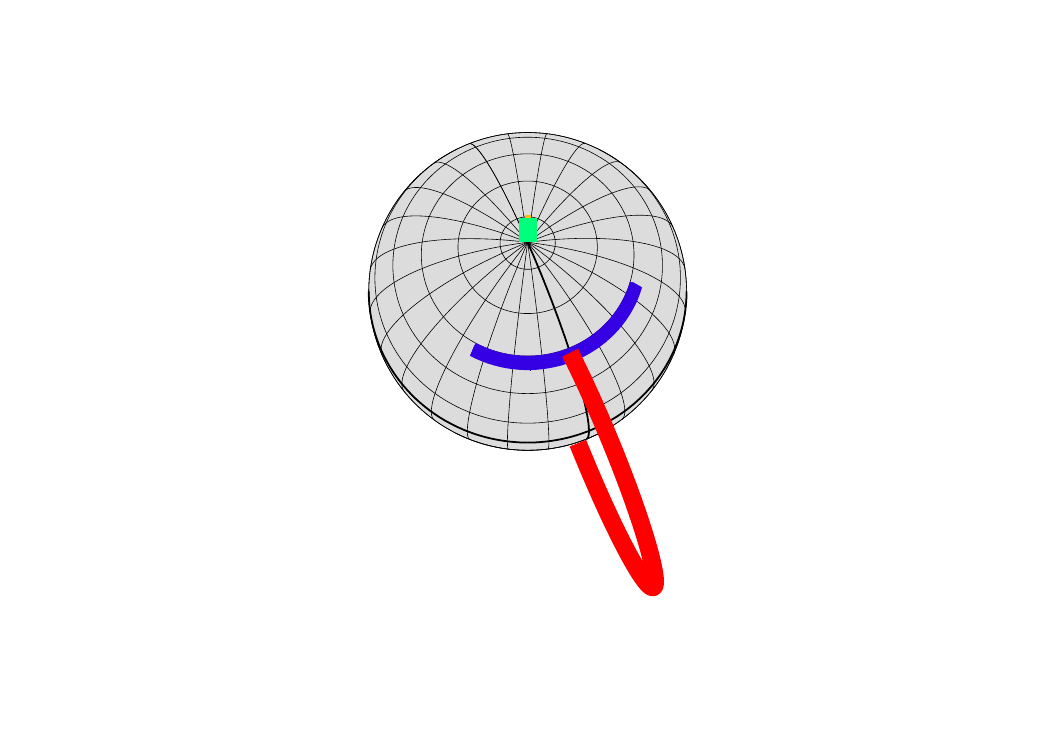}\includegraphics[width=4.3 cm, trim=150 70 140 0]{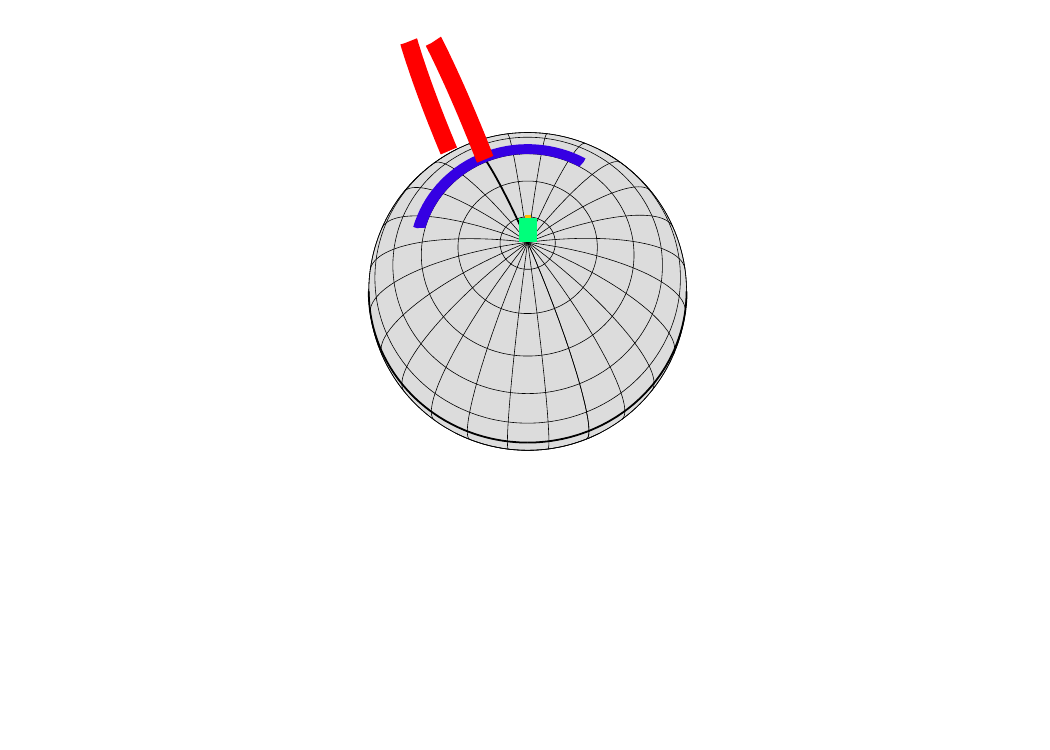}
\caption{(Top) A model for the circularly polarized flux from the PSR of V1082~Sgr (blue line). Black points represent the observations. (Middle) 2D maps of the total intensity and circularly polarized flux in 12 spin phases separated by 30\degr. The flux scale increases from black to white, passing through orange. (Bottom) A projected view of the PSR region (in blue) over the WD surface in two phases of the WD rotation separated by half a cycle. The magnetic axis is shown as a green line and the magnetic field line that crosses the center of the PSR is shown as a red line. See Section~\ref{sec:cyclops_model} for details of the model.}
\label{fig:cyclops_model}
\end{figure}

The top panel of Figure~\ref{fig:cyclops_model} presents the model superposed on \sgr\ observations. The model parameters are presented in Table~\ref{tab:model}. The reader can find a proper explanation of all parameters in \citet{Costa_2009} and \citet{Belloni_2021_cyclops}. The values of WD mass, the mass-accretion rate, and the inclination are considered fixed according to \citet{2019MNRAS.489.3031X}, which provides the most updated estimates. Using the values of \citet{Tovmassian_2018b} and \citet{Bernardini_2013} to those parameters, we obtain a similar curve and the overall properties of the PSR are the same. For example, the magnetic field strength of the best fit model changes to $9.4~\times~10^{6}$~G.
The {\sc cyclops} code considers a centered dipolar magnetic field, whose axis can point to any direction. But, to minimize the number of free parameters, we adopted a magnetic axis parallel to the WD rotation axis. The model provides a WD magnetic field strength of approximately 10$^{7}$~G, a value within the standard range of IPs. The temperature in the modeled PSR ranges from 1.2 to 17.1~keV in the base of the PSR and in the shock front, respectively. This is consistent with the fit of the \xmm\ and \swift\ spectra presented by \citet{Bernardini_2013}, who uses two {\sc MEKAL} components of 0.12~keV and 13.2~keV. Using the same data but a different modeling, \citet{2019MNRAS.489.3031X} estimated a $T_{max}$ of $35^{+14}_{-9}$~keV, which is considerably larger than our results, but yet barely consistent considering the different approaches of the modeling.

In a dipolar magnetic geometry (as adopted by {\sc cyclops}), there can be two PSRs: each of them associated with one magnetic hemisphere. Our V1082 Sgr model adopts a single PSR region, and this does not imply any approximation because a possible second PSR associated with the magnetic pole pointing backward is always hidden from the observer view (see Figure~\ref{fig:cyclops_model}, bottom panel). This is mainly a consequence of the small inclination of the system, but it also results from the angle between the WD rotation axis and the PSR. Therefore, a second PSR cannot contribute to the observed flux, even if two accretion columns exist.

To fit the observed shape of the polarized flux phase diagram, it is necessary to consider an azimuthally elongated PSR: the emission map of the PSR is presented in the middle panel of Figure~\ref{fig:cyclops_model}. As the PSR footprint reflects the threading region (i.e., the region in the equatorial plane from where the accretion flow begins to follow the WD magnetic field lines) geometry, the coupling of the flux with the magnetic field lines should occur along an extended region in the equatorial plane. 
Moreover, this PSR is observable in all spin phases, so no modulation due to self-eclipse by the WD is possible in this model. The changing view of the PSR causes the variation of the optical circular polarization, since the cyclotron emission depends on the angle between the magnetic field and the observer.

The {\sc cyclops} code considers the entire geometry of the magnetic accretion column.
In particular, the model geometry is built from the threading region. Our model leads to a threading region located at 2.1 WD radii (R$_{\rm WD}$) from the WD surface. Using the model parameters, the magnetospheric radius of \sgr\ is 3.2~R$_{\rm WD}$ (considering the expression of \citealt{ferrario1989}), which is similar to the threading region value of our model providing a consistent picture to the system configuration.

\begin{deluxetable*}{lll}
%\tablenum{1}
\tablecaption{Parameters of the PSR of \sgr\ modeled using the {\sc cyclops} code. See text for details.\label{tab:model}}
%\tablewidth{0pt}
\tablehead{
\colhead{Parameter} & \colhead{Value} & \colhead{Source}  
}
\startdata
%\noalign{\smallskip}
\hline \noalign{\smallskip} 
\multicolumn{3}{c}{Fixed parameters} \\
\hline\noalign{\smallskip} 
Inclination & 18$\degr$ & \cite{2019MNRAS.489.3031X} \\
WD mass & 0.77~M$_\odot$ & \cite{2019MNRAS.489.3031X} \\
Mass-transfer rate & $1.2~\times 10^{-9}$~M$_\odot$ yr$^{-1}$ & \cite{2019MNRAS.489.3031X} \\
\hline\noalign{\smallskip} 
\multicolumn{3}{c}{Free parameters} \\
\hline\noalign{\smallskip} 
Angle between the center of the PSR and the WD rotations axis & 44$\degr$ & From the fit \\
Magnetic field in the magnetic pole & $11~\times~10^{6}$ G & From the fit \\
Longitudinal extension of the threading region & 103$\degr$ & From the fit \\
Radial extension of the threading region & 0.16 R$_{\rm WD}$ & From the fit \\
\hline\noalign{\smallskip} 
\multicolumn{3}{c}{Some quantities obtained from the model}\\
\hline\noalign{\smallskip} 
Shock temperature & 17.1 keV  & Calculated from the model parameters \\
Threading radius & 2.1 R$_{\rm WD}$ & Calculated from the model parameters \\
PSR height & 0.023 R$_{\rm WD}$ & Calculated from the model parameters \\
Latitudinal extension of the PSR & 4.3$\degr$ & Calculated from the model parameters \\
\hline\noalign{\smallskip} 
\enddata
\tablecomments{R$_{\rm WD}$: white-dwarf radius.}
\end{deluxetable*}

\section{Discussion} \label{discussion}

\cite{2016ApJ...819...75T} and \cite{Tovmassian_2018a, Tovmassian_2018b} argued that \sgr\ could be a prepolar. 
The arguments in favor of the prepolar hypothesis for V1082\,Sgr are the very long \porb, at which it is difficult to form a Roche-lobe filling CV with a K-type secondary, and the very irregular, fast variability, associated with the changing accretion rate. The counter-argument is that even a fast-rotating, magnetically active K-star could hardly provide stellar winds of the order of $10^{-9}-10^{-10}~\rm{M}_{\odot}~\rm{yr}^{-1}$ needed to explain the high luminosity, larger than $10^{34}$~erg\,s$^{-1}$, of V1082\,Sgr.
\citet{2019MNRAS.489.3031X} proposed an alternative scenario in which the system underwent a thermal time-scale mass-transfer episode in the past and currently is a Roche-lobe filling IP with a low-mass secondary, which, however, has a temperature corresponding to an early K-type star. 
The circular polarization in magnetic CVs comes from a structure anchored on the WD, so the period of 1.943~h found in \sgr\ is certainly the WD \pspin. It rules out a polar or a prepolar classification since these are synchronous systems. Therefore, \sgr\ should be a long orbital-period IP in line with the scenario proposed by \citet{2019MNRAS.489.3031X}. 

Unlike the polars, which have large and ubiquitous circular polarization levels of several tens of percent (e.g., \citealt{Oliveira_2019}), IPs have low level circular polarization or even non-detected polarized flux. For example, V2731~Oph \citep{Butters_2009} is one of the most polarized IPs: the spin-folded circular polarization curve shows a peak-to-peak amplitude of 8.26~$\pm$~1.56\% in B band (and about 3\% in U, V, R and I bands). For V1082~Sgr, we found an amplitude for the circular polarization of around 1.0\%, in agreement with typical values found in other IPs. 

\begin{figure}[tbh]
\includegraphics[width=\columnwidth]{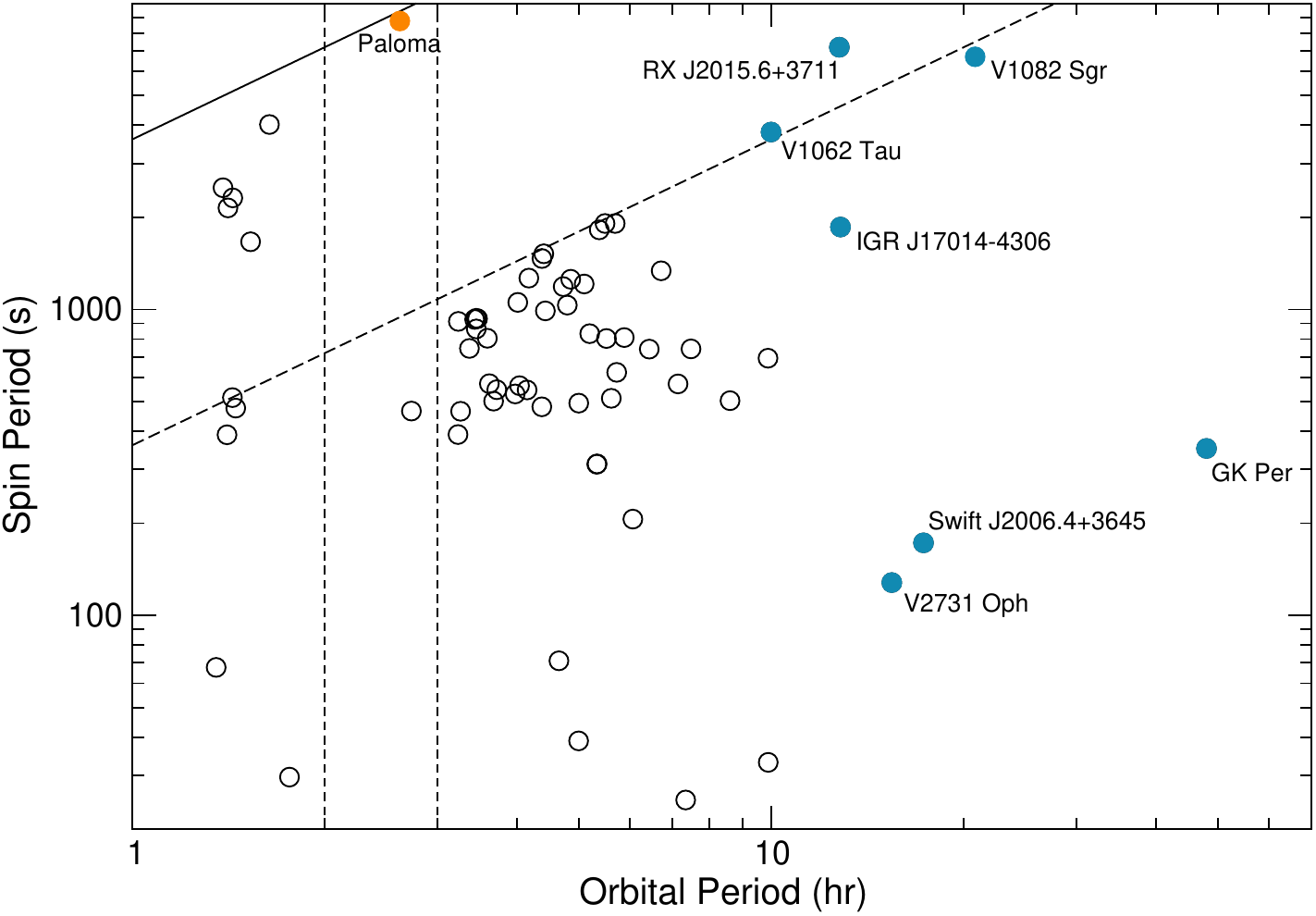}
\caption{\pspin versus \porb\ for IPs. V1082~Sgr and other long \porb\ objects discussed in the text are highlighted in blue, while the nearly synchronous, short \porb\ IP Paloma is depicted in orange. The vertical lines are the period gap interval of 2 to 3~h and the diagonal lines represent \pspin~=~\porb\ (solid) and \pspin ~= 0.1~$\times$~\porb (dashed). Data, except for V1082 Sgr, from Mukai's IP Homepage (see text).\label{fig:psporb}}
\end{figure}

\sgr\ has \pspin /\porb\ equal to 0.093, which is a typical value for an IP \citep{King_1991, Mukai_2017}. Moreover, that \pspin\ makes V1082\,Sgr the IP with the third longest \pspin, after RX~J2015.6+3711 \citep{Coti_2016} and Paloma (\citealt{Schwarz_2007,2016ApJ...830...56J,2017RNAAS...1...29T}; \citealt {2023AJ....165...43L}) with 2~h and 2.27~h, respectively. Paloma has a value of \pspin /\porb\  larger than 0.85, which is very odd for an IP. It is a member of a (new) group of several magnetic CVs whose WDs rotate with frequencies around 90\% of the orbital one \citep{2024MNRAS.527..774P}. On the other hand, RX~J2015.6+3711 is also a long orbital-period IP, with a period of around 12.76~h \citep{Halpern_2018}. Table~\ref{tab:long-period} enumerates the IPs having the longest ($\gtrsim 10$~h) \porb. 
Figure~\ref{fig:psporb} shows the distribution of currently confirmed IPs in the \pspin\ versus \porb\ plane from Mukai's Intermediate Polars Homepage\footnote{\url{https://asd.gsfc.nasa.gov/Koji.Mukai/iphome/iphome.html}}. We added to this plot the location of \sgr, which is close to the \pspin ~=~0.1~$\times$~\porb\ line and in the upper right limit of this distribution. We also highlighted in this plot other long orbital-period IPs. 

The accretion mode/geometry of IPs is an old and complex subject \citep[see for instance,][]{1991MNRAS.251..693H}. \cite{Norton_2004} and \cite{Norton_2008} shed some light on this topic through simulations of accretion onto magnetic white dwarfs. These studies showed numerically that \pspin /\porb\ can be used to unravel the accretion modes of IPs. We can have a hint whether \sgr\ is in spin equilibrium, understood as the balance of angular momentum in the accretion process, by checking its position in Figure~2 of \citet{Norton_2004}. This figure shows the results of simulations for \porb\ from 80~min to 10~h, so some extrapolation should be considered for \porb\ = 20~h: the model line for such a period should be to the right of the models lines depicted in the figure. Considering a magnetic moment of $4.3 \times 10^{33}$~$\rm{G\,cm^{3}}$, \sgr\ would be placed above the spin equilibrium line. In case it is true, the WD should be in a spin-up state, i.e., evolving to shorter \pspin\ to achieve the spin equilibrium. 
The corotation radius, $R_{co}$, of \sgr\ for a Keplerian period equals \pspin\ is 69~$R_{WD}$. 
$R_{co}$ is not exactly the equilibrium radius for the magnetic accretion, but they are assumed to be similar \citep[see][and references therein]{Norton_2004}. The circularization radius, $R_{circ}$, of \sgr\ is around 40~$R_{WD}$, so much smaller than the $R_{co}$. These numbers are consistent with \sgr\ not being a synchronous system, situation in which $R_{mag}$ should also be larger or around $R_{circ}$. Our estimate of $R_{mag}$ is approximately 3~$R_{WD}$, which is consistent with this scenario. Moreover, the huge difference between $R_{mag}$ and $R_{co}$ also points to a system out of equilibrium.

An evaluation of the accretion geometry of \sgr\ can be done based on the results of \citet{Norton_2008}.
Their Fig.~2 illustrates the possible accretion structures for an asynchronous magnetic CV as a function of \pspin\ and magnetic field for \porb~=~4~h and a mass ratio ($q =M_{2}/M_{1}$) equal to 0.5. 
The central panel of this figure shows that the stream-like accretion can occur for any value of magnetic moment.
In particular, a stream-like geometry (i.e., with no disk at all - top left panel of that figure) can occur for small values of the magnetic field, if \pspin\ is long enough. In this case, the threading region is located near the WD. 
To verify if this could be the case for \sgr, we should inspect \citet{Norton_2008}'s Figure~5. 
For $q = 0.5$, 
values of \pspin / \porb~$<$ 0.1 correspond to disk-like accretion. \sgr\ has \pspin /\porb\ ratio of 0.093 and an estimated mass ratio of around 0.7 \citep{2019MNRAS.489.3031X}: this mass ratio decreases the aforementioned limit. So this system would be in the limit between a disk-like and stream-fed accretion. 
Therefore, we could suppose that
\sgr\ is located in a region where stream accretion is possible. A stream geometry is more prone to strong changes in the accretion states because there is no disk to act as a material reservoir. 
For example, the substantial alterations in brightness observed in polars are likely due to this phenomenon. Therefore, it is possible that the variability of the accretion rates of \sgr\ is related to the absence of an accretion disk in the system. 
Moreover, a stream-like geometry is consistent with a spin-up WD \citep{Norton_2008}, as discussed in the previous paragraph.
However, the conclusion that \sgr\ is a stream-fed system should be taken with a grain of salt because $q$ can be as small as 0.25 (Tovmassian, G. et al., in preparation), placing \sgr\ in the disk-accretion region.

\cite{Bernardini_2013} reported an extensive study of X-ray data of \sgr\ obtained with \xmm, \suzaku, and \swift~ from 2008 until 2012. The light curve of \xmm-PN data, with energies ranging between 0.3 and 15~keV, shows a modulation with a period of about 2 hours, which was only present during some observation intervals. This modulation is energy-dependent, with a pulsed fraction decreasing towards higher energies.
A similar, $\sim$2-h modulation was also detected during a \suzaku\ observation and less significantly during \swift\ pointings. This 2-h modulation was not persistent, so it was interpreted as due to non-coherent transient accretion.

We argue that the intermittent 2-h modulation seen in X-rays is also related to the WD spin. Important radiative cooling processes in the PSR of magnetic CVs are the cyclotron emission, responsible for the optical polarization, and bremsstrahlung, which contributes with an important fraction of X-ray emission. While the cyclotron emission is intrinsically anisotropic and, therefore, can be modulated with the WD \pspin, the bremsstrahlung emission is usually optically thin and isotropic, so there should be external mechanisms to produce flux variability. In magnetic CVs in general and in IPs in particular, the modulation of the X-ray emission can be produced by the self-occultation of PSR by the WD or by phase-dependent absorption/scattering of the PSR emission by structures lying between the PSR and the observer, as the upper portion of the magnetic column \citep{Norton_1993, Belloni_2021_cyclops}. 
Self occultation is a geometric effect and does not depend on energy.
On the other hand, photoelectric absorption depends on energy, decreasing from soft to hard X-rays, consistent with the pulsed-fraction energy dependence reported by \citet{Bernardini_2013}. Moreover, the inclination of \sgr\ and the proposed location of its PSR prevent any modulation due to self eclipse. However, the intermittence of the X-ray modulation remains to be explained. It could be connected with the extreme variation of the accretion rate, which can occasionally drop to negligible values in \sgr\  \citep{2016ApJ...819...75T}. The accretion rate affects not only the density of the PSR itself but also the density of the ``upper'' structures responsible for the absorption. In particular, a stream-accretion geometry is more susceptible to reflect the mass-accretion variations.

\,

\begin{deluxetable*}{ccccc}[tbh]
\tablecaption{IPs with \porb\ longer than $\sim$10~h.\label{tab:long-period}}
\tablewidth{0pt}
\tablehead{
\colhead{Object} & \colhead{\porb}  &
\colhead{\pspin} & \colhead{\pspin /\porb} & \colhead{Circular Polarization}\\
& \colhead{(h)} & \colhead{(h)}  & \colhead{(h)} &  \colhead{(\%)}
}
%\decimalcolnumbers
\startdata
GK~Per & 47.92 & 0.097 & 0.002  & +0.03~$\pm$~0.10 (1) \\
V1082~Sgr & 20.82 & 1.943 (2) & 0.093 & 0 to 1 (2)  \\
Swift J2006.4+3645 & 17.28 & 0.048 & 0.003 & $-$\\
V2731 Oph & 15.42 & 0.035 & 0.002 & -4 to +4.26 (3)  \\
IGR J17014-4306 & 12.82 & 0.516 & 0.040 & 0 to -2.5 (4)\\
RX J2015.6+3711 & 12.76 & 1.998 & 0.156 & $-$ \\
V1062 Tau & 9.98 & 1.056 & 0.106 & $-$\\
\enddata
\tablecomments{The values of \porb\ and \pspin\ are from the Mukai's IP Homepage and references therein.\\
(1) \cite{Stockman_1992}, reported the mean value.\\
(2) This paper.\\
(3) \cite{Butters_2009}, for B band.\\
(4) \cite{Potter_2018}.\\
\\
\\
}
%}
\end{deluxetable*}

\section{Conclusions}
\label{sec:conclusions}

We have detected a periodicity of 1.943~$\pm$~0.002~h in the circular polarization of \sgr, with an amplitude smaller than 1\%. We interpret this period as the rotation of a magnetic WD, since the polarized flux is produced by cyclotron emission from the post-shock region near the WD surface. This indicates that \sgr\ is an IP, in agreement with the suggestion of \citet{2019MNRAS.489.3031X}. Also, the intermittent period near 2~h found in X-ray data by \cite{Bernardini_2013} is possibly related to the WD spin and could be associated with phase-dependent absorption structures in the accretion column.

The {\sc cyclops} model of the cyclotron emission of V1082\,Sgr yields a magnetic field strength of 10$^{7}$~G and temperatures that vary from 17~keV in the top of the PSR to 1~keV in its bottom, in line with X-ray determinations by \citet{Bernardini_2013}. According to the model, the threading region is placed at a distance of 2.1~R$_{\rm WD}$ from the WD surface, roughly at a similar position to the estimated magnetospheric radius. Considerations about this radius and the expected corotation radius indicate that V1082\,Sgr is out of spin equilibrium. This conclusion is reinforced by a comparison of the values of the \pspin /\porb\ ratio and the estimated magnetic moment with published simulations, which indicate that the WD is in a spin-up state. In spite of the reasonable {\sc cyclops} fitting, more restrictive modeling of the PSR should be based on multi-band emission since the cyclotron emission is highly dependent on the frequency. The \pspin\ versus \porb\ relation also points to a possible stream accretion mode on V1082\,Sgr, which in turn could be associated with the recorded highly variable optical brightness states and intermittent X-ray periodic modulation due to the absence of a disk as a mass reservoir. 

V1082\,Sgr is, therefore, the known IP with the second longest \porb\ and with the third longest \pspin, which places it as an important case to the understanding of the evolution of magnetic cataclysmic variables. 
We foresee many additional studies for this systems as, for instance: an observational verification of the WD spin-up; a better constrained model for its PSR; and Doppler tomography to better define the accretion geometry.

\begin{acknowledgments}

IJL acknowledges support from grant ANPCYT-PICT 0901/2017 (Argentina) and \textit{Coordenação de Aperfeiçoamento de Pessoal de Nível Superior} -- Brazil (CAPES, number \#88887.913793/2023-00).
GT was supported by grants IN109723 and IN110619 from the  \textit{Programa de Apoyo a Proyectos de Investigación e Innovación Tecnológica} (PAPIIT).
This research was supported in part by grant NSF PHY-1748958 to the Kavli Institute for Theoretical Physics (KITP).
CVR thanks the Brazilian Ministry of Science, Technology and Innovation (MCTI) and the Brazilian Space Agency (AEB) by the support throught the PO 20VB.0009 and also the \textit{Conselho Nacional de Desenvolvimento Científico e Tecnológico} (CNPq Grant: 310930/2021-9).
This research was supported in part by the National Science Foundation under Grant No. NSF PHY-1748958 (CVR).
%Alexandre
ASO acknowledges \textit{São Paulo Research Foundation} (FAPESP) for financial support under grant \#2017/20309-7.
%Juan 
GJML is member of the CIC-CONICET (Argentina). 
ACM thanks the CNPq (Proc: \#150737/2024-6). DCS thanks the CAPES (Proc: \#88887.714650/2022-00 and \#88881.846501/2023-01). 
WS thanks the CNPq (Proc: \#318052/2021-0, \#300343/2022-1, \#300834/2023-3, \#301366/2023-3, \#300252/2024-2 and \#301472/2024-6).
FF thanks the CNPq (Proc: \#141350/2023-7). 
\end{acknowledgments}


\vspace{5mm}
\facilities{LNA:0.6m.}


\software{\mbox{IRAF} \citep{Tody/1986, Tody/1993}, \mbox{{\sc PCCDPACK\_INPE}} \citep{2018Pereyra, pereyra/2000}, \mbox{{\it PERANSO}} \citep{Paunzen_2016}.
          }



\appendix

\section{Observations of polarimetric standard stars} \label{pol_standard}

In this appendix, we present a summary of the measurements of polarimetric standard stars obtained in the same observational run of \sgr. Table~\ref{pol_stard} shows the average linear and circular polarizations of linearly polarized stars. The position angle values are not calibrated to the equatorial system. The circular polarization values are consistent with zero. Table~\ref{unpol_stard} shows the results for unpolarized stars. Again the values of circular polarization are consistent with zero. These results demonstrate that our measurements are not affected by instrumental circular polarization in a level of 0.1\% or smaller. 

\begin{deluxetable*}{ccccc} [ht!]
\tablecaption{Average polarization measurements of linearly polarized standard stars.\label{pol_stard}}
%\tablewidth{0pt}
\tablenum{A1}
\tablehead{
\colhead{Object} & \colhead{$P_{l}$} & \colhead{PA}  &
\colhead{$P_{C}$} & \colhead{Number of} \\
& \colhead{(\%)} & \colhead{(deg)} & \colhead{(\%)}  &  observations
}
%\decimalcolnumbers
\startdata
HD~161056 & 3.64$~\pm~$0.11 & 46.6 & 0.08$~\pm~$0.07 & 6 \\
Hilt~785 & 4.08$~\pm~$0.05 & 154.2 & 0.03$~\pm~$0.03 & 5\\
Hilt~781 & 3.81$~\pm~$0.05 & 152.3 & 0.02$~\pm~$0.03 & 6\\
\enddata
%\tablecomments{}
\end{deluxetable*}

\begin{deluxetable*}{cccc}[ht!]
\tablecaption{Average polarization measurements of unpolarized standard stars.\label{unpol_stard}}
%\tablewidth{0pt}
\tablenum{A2}
\tablehead{
\colhead{Object} & \colhead{$P_{l}$}  &
\colhead{$P_{C}$} & \colhead{Number of} \\
& \colhead{(\%)} & \colhead{(\%)}  &  observations
}
%\decimalcolnumbers
\startdata
WD~2007-303 & 0.10$~\pm~$0.11 & 0.12$~\pm~$0.17  & 4\\ 
WD~2149+021 & 0.13$~\pm~$0.19 & 0.07$~\pm~$0.11 & 2 \\
WD~2039-202 & 2.22$~\pm~$0.56 & 0.09$~\pm~$0.44 & 2\\
\enddata
%\tablecomments{}
\end{deluxetable*}


\bibliography{sample631}{}
\bibliographystyle{aasjournal}



\end{document}